\documentclass[12pt]{amsart}

\usepackage{amsmath,amssymb,amsthm,a4,bm}

\newcommand{\ie}{i.\,e.}

\newtheorem{theorem}{Theorem}
\newtheorem{proposition}{Proposition}
\newtheorem{lemma}{Lemma}

\newtheorem*{remark*}{Remark}
\newtheorem*{remarks*}{Remarks}

\newcommand{\allspace}{{\mathbb Z}^d}
\newcommand{\cH}{{\mathcal H}}
\newcommand{\cU}{{\mathcal U}}
\newcommand{\cD}{{\mathcal D}}
\newcommand{\cC}{{\mathcal C}}
\newcommand{\sP}{{\mathsf P}}
\newcommand{\cS}{{\mathcal S}}
\newcommand{\cM}{{\mathcal M}}
\newcommand{\unit}{{\mathbf 1}}
\newcommand{\tA}{\tilde{\alpha}}
\newcommand{\tB}{\tilde{\beta}}

\newcommand{\oA}{\overline{A}}

\begin{document}

\title[Fluctuations of Observables in NESS]{{\bf On the Fluctuations of Macroscopic Observables in Quantum Nonequilibrium Steady States}}

\author{Walid K. Abou Salem}

\address{Walid K. Abou Salem, Department of Mathematics, Univesity of Toronto, Toronto, Ontario, Canada M5S 2E4}

\email{walid@math.utoronto.ca}


\begin{abstract}

The fluctuations of macroscopic observables in quantum systems which are in a nonequilibrium steady state are studied rigorously in the thermodynamic limit. In particular, the nonequilibrium steady state (NESS) of a quantum spin system that is composed of two infinite reservoirs at different temperatures coupled in a bounded region is considered. Under suitable assumptions on the interaction and the asymptotics of the fluctuations, a nonequilibrium central limit theorem for the fluctuations of the NESS expectation value of the empirical average of observables is proven by showing the analyticity of the corresponding moment generating function in a neighbourhood of the origin using the cluster expansion method. Furthermore, the large deviation principle is proven for a class of observables by establishing the existence and differentiability of the corresponding moment generating function. 

\end{abstract}


\maketitle
\begin{center}
{\em Dedicated to J\"urg Fr\"ohlich on the occasion of his sixtieth birthday.}
\end{center}
\bigskip


\section{Introduction}

The central limit theorem and large deviation theory have a long history in equilibrium statistical mechanics, where they play an essential role in understanding the fluctuations of macroscopic observables in systems at equilibrium (;see for example \cite{Ellis05}). For instance, it is well-known that the relative entropy density governs the large deviations of empirical measures for Gibbs random fields in classical statistical mechanics, see for example \cite{Ellis05} and \cite{DZ98}. Noncommutative central limit theorems have also been established for the fluctuations of macroscopic observables in equilibrium quantum statistical mechanics.\cite{GVV1,GVV2,M1,M2} Recently, the large deviation principle has been proven for fluctuations of some observables of quantum systems at equilibrium. For continuous quantum systems, the large deviation principle (and a central limit theorem) for the fluctuations of the particle density has been established for ideal quantum gases in \cite{LLS00}, and dilute quantum gases in \cite{GLM02}, using the cluster expansion method. Furthermore, the large deviation principle and a central limit theorem has been shown to hold for fluctuations of onsite observables in quantum spin systems at equilibrium in \cite{NR04} using a high temperature cluster expansion, while (part of) the large deviation principle has been proven for a general class of observables of quantum spin systems at equilibrium in \cite{LR-B05} using the Gibbs condition.  

Unlike in systems at equilibrium, fluctuations in nonequilibrium steady states (NESS) is poorly understood. Recently, there has been encouraging progress in understanding fluctuations of observables in specific classical statistical systems at NESS in the {\it hydrodynamic limit}, particularly in simple exclusion processes, \cite{DLS1,DLS2}, and in stochastic Hamiltonian systems, \cite{BDGJL02,BDGJL03}, based on ideas from large deviation theory. There is no analogue of the latter results in nonequilibrium quantum statistical mechanics starting from a (time-reversal invariant) microscopic description of quantum systems. \footnote{Fluctuations for stochastic quantum systems out of equilibrium has been studied in \cite{Sewell06} using a phenomenological approach.} 

In this note, we consider the nonequilibrium steady state (NESS) of a simple, yet nontrivial, quantum system, a quantum spin system consisting of two infinite reservoirs at different temperatures which are coupled in a bounded region, and we study in a mathematically rigorous manner the fluctuations of the NESS expectation value of the empirical average of observables in the {\em thermodynamic limit}. In particular, under suitable assumptions on the decay of interactions and the asymptotic evolution of the fluctuations, we prove the existence and analyticity of the corresponding moment generating function in a neighbourhood of the origin. By Proposition 1 in \cite{Bryc93}, this implies that the fluctuations satisfy a nonequilibrium central limit theorem. In the special case when there is no coupling between the reservoirs, the state of each reservoir is an equilibrium KMS state, and we recover a central limit theorem for the fluctuations of general observables in quantum spin systems at equilibrium, which was mentioned in \cite{NR04} and announced in \cite{LR-B05}, but which has not been explicitly proven so far. Furthermore, the differentiability of the moment generating function for a class of observables implies the large deviation principle by the Ellis-G\"artner theorem, \cite{DZ98}. 

The organization of this note is as follows. In Section 2, we discuss the model system we consider, recall the definition of NESS, and describe the main problem. In Section 3, we state the main result, which we prove in section 4. Our analysis relies on the algebraic formulation of quantum statistical mechanics, \cite{BR, Ruelle}, the notion of nonequilibrium steady states, \cite{Ruelle1,Ruelle2}, large deviation theory, \cite{DZ98}, the Kotecky-Preiss criteria for the convergence of the cluster expansion, \cite{KP86,Ueltschi04}, and the analysis in \cite{NR04,LR-B05} of fluctuations of observables in quantum spin systems at equilibrium. In order to get spatial (and temporal) profiles and to make contact with the quantum version of the results in \cite{DLS1,DLS2,BDGJL02,BDGJL03}, one needs to take the {\it hydrodynamic limit} of the reservoirs rather than the {\it thermodynamic limit}. Further physical applications and a discussion of the hydrodynamic limit are the subject of future work. We note that combining the results  of \cite{FMU03} and \cite{LLS00}, one can easily extend the analysis of this note to continuous fermionic quasi-free quantum systems in NESS.


\subsection{Notation}

Consider a $d$-dimensional lattice ${\mathbb Z}^d.$ For $a=(a_1, \cdots, a_d)\in {\mathbb Z}^d,$ such that $a_i>0, i=1,\cdots, d,$ we define $\Lambda_a\subset {\mathbb Z}^d$ as the parallelepiped   
\begin{equation*}
\Lambda_a:= \{ x=(x_1,\cdots,x_d) \in {\mathbb Z}^d : 0\le x_i < a_i, i=1,\cdots, d\} .
\end{equation*} 
The set of translates of $\Lambda_a$ by $na:= (n_1 a_1, \cdots , n_d a_d) ,n_i\in {\mathbb Z}, i=1,\cdots, d,$ form a partition ${\mathbf P}_a$ of $\allspace.$ For $\Lambda\subset \allspace,$ we let $\overline{N}_\Lambda(a)$ be the number of sets of this partition which have a nonempty intersection with $\Lambda,$ and we let $N^0_\Lambda(a)$ be the number of sets of this partition which are included in $\Lambda.$ Finite subsets $\Lambda\subset \allspace$ are defined to tend to $\allspace$ in the sense of van Hove, $\Lambda \nearrow \allspace,$ if 
\begin{equation*}
\lim \frac{\overline{N}_\Lambda (a)}{N^0_\Lambda(a)} =  1 
\end{equation*}
for all partitions ${\mathbf P}_a.$\cite{BR, Ruelle}


\section{The Model and Description of the Problem}\label{sec:model}

\subsection{The Model}

We consider a quantum spin system composed of two infinite reservoirs, ${\mathbf R}_{i}, i=1,2,$ at different temperatures which are coupled in a bounded region $\Lambda_0.$\footnote{We refer the reader to \cite{BR} for a detailed exposition of quantum spin systems.} For the sake of concreteness, we assume that each reservoir,  ${\mathbf R}_1$ and ${\mathbf R}_2,$ occupies ${\mathbb Z}^d.$ 

We associate with every $x\in \allspace$ two finite dimensional complex Hilbert spaces $\cH_x^1$ and $\cH_x^2.$ Furthermore, for $X\subset \allspace$ we define the local Hilbert spaces
\begin{eqnarray*}
\cH_X^1 := \otimes_{x\in X}{\cH_x^1} , \\
\cH_x^2 := \otimes_{x\in X}{\cH_x^2}, \\
\cH_X:= \cH_X^1\otimes\cH_X^2.
\end{eqnarray*}
We denote by $\cU_X^i,i=1,2,$ the kinematical algebra of local observables, the $C^*$-algebra defined by $\cU_X^i={\mathcal B}(\cH_X^i),$ the algebra of bounded operators on $\cH_X^i, i=1,2.$ If $X\cap X'=\phi,$ then $\cH^i_{X\cap X'}=\cH^i_{X}\otimes\cH^i_{X'},i=1,2,$ and $\cU^i_{X}$ is isomorphic to  $\cU^i_{X}\otimes\unit^i_{X'},$ where $\unit^i_{X'}$ is the identity operator on $\cH^i_{X'}, i=1,2.$ Note that, by construction, $\cU^i_{X}\subseteq \cU^i_{X'}$ if $X\subseteq X',$ and $[\cU^i_{X},\cU^i_{X'}]=0$ whenever $X\cap X'=\phi,$ where $[\cdot, \cdot]$ stands for the commutator. Furthermore, $[\cU^1_{X}, \cU^2_{X'}]=0$ for every $X,X'\in {\mathbb Z}^d.$ We denote by $\cU_X:=\cU_X^1\otimes\cU^2_X.$
The quasi-local $C^*$-algebras $\cU,\cU_{R_i}, i=1,2,$ are given by
\begin{equation*}
\cU=\overline{\cup_{X,X'\subset\allspace}\cU^1_X\otimes\cU^2_{X'}}, \cU_{R_i}=\overline{\cup_{X\subset \allspace}\cU^i_X}, i=1,2,
\end{equation*}
where $\overline{(\cdot)}$ denotes the norm closure. The quasi-local $C^*$-algebras $\cU$ and $\cU_{R_i}$ have a common identity element $\unit.$

In order to specify the dynamics of the uncoupled system, we introduce the interaction $\Phi^i$ in each reservoir ${\mathbf R}_i, i=1,2,$ which is defined on finite subsets of $\allspace.$ For $\Lambda\subset {\mathbb Z}^d,$ such that $|\Lambda|<\infty,$
\begin{equation*}
\Phi^i: \Lambda\rightarrow \cU^i_\Lambda, 
\end{equation*}
such that $\Phi^i(\Lambda)$ is a selfadjoint element of $\cU^i_\Lambda.$
We define the local Hamiltonians
\begin{equation*}
H_{R_i}(\Lambda):= \sum_{X\subset \Lambda} \Phi^i (X),i=1,2 .
\end{equation*}
We also introduce the interaction $\Psi$ corresponding to the local coupling of the reservoirs,
\begin{equation*}
\Psi: {\mathbb Z}^d\supset\Lambda\rightarrow \cU_\Lambda,
\end{equation*}
such that $\Psi(\Lambda)$ is a selfadjoint element of $\cU_\Lambda$ and $\Psi(\Lambda)=0$ if $\Lambda\cap\Lambda_0=\phi.$ The interaction Hamiltonian corresponding to the local coupling is formally given by 
\begin{equation*}
V=\sum_{\stackrel{\Lambda\subset\allspace}{\Lambda\cap\Lambda_0\ne \phi}} \Psi(\Lambda).
\end{equation*}
We make the following assumptions. 

\begin{itemize}

\item[(A1)] {\it Decay of the interaction.} $\exists\lambda>0$ such that 
\begin{equation*}
\|\Phi^i\|_\lambda:= \sup_{x\in\allspace}\sum_{n\ge 0} e^{n\lambda} \sum_{\stackrel{X\ni x}{|X|=n+1}}\|\Phi^i(X)\| <\infty , \; i=1,2.
\end{equation*}

\item[(A2)] {\it Decay of the coupling.} 
\begin{equation*}
\|\Psi\|:= \sup_{x\in\allspace} \sum_{\stackrel{X\subset\allspace}{X\ni x}}\|\Psi(X)\| <\infty .
\end{equation*}
\end{itemize}

It follows from assumption (A2) that the interaction Hamiltonian $V$ is bound,
\begin{equation*}
\|V\| \le |\Lambda_0| \|\Psi\| <\infty.
\end{equation*} 
We introduce the linear operators $\delta$ and $\delta_0$ with domain $\cD=\cup_{\Lambda\subset\allspace}\cU_\Lambda ,$ such that, for $A\in \cU_{\Lambda'} , \Lambda'\subset \allspace,$
\begin{equation*}
\delta (A):= i \sum_{i=1}^2 \sum_{\stackrel{X\subset\allspace}{X\cap\Lambda' \ne \phi}} [\Phi^i(X),A] + i[V,A],
\end{equation*}
and 
\begin{equation*}
\delta_0(A):= i \sum_{i=1}^2 \sum_{\stackrel{X\subset \allspace}{X\cap\Lambda' \ne \phi}} [\Phi^i(X),A].
\end{equation*}
We also introduce, for $\Lambda\subset {\mathbb Z}^d,$ the linear operators $\delta_\Lambda$ and $\delta_{0,\Lambda}$ which are given by
\begin{equation*}
\delta_\Lambda (A):= i\sum_{i=1}^2[H_{R_i}(\Lambda),A] + i[V,A],
\end{equation*}
and 
\begin{equation*}
\delta_{0,\Lambda} (A) := i\sum_{i=1}^2 [H_{R_i}(\Lambda),A],
\end{equation*}
for $A\in \cD.$ The following lemma follows from assumptions (A1) and (A2) and Theorem 6.2.4 in \cite{BR}. 

\begin{lemma}\label{lm:Evolution} Suppose assumptions (A1) and (A2) hold. Then $\cD$ is a norm-dense $*$-subalgebra of analytic elements of the (norm) closures $\overline{\delta}$ and $\overline{\delta_0}$ of $\delta$ and $\delta_0$ respectively. Furthermore, $\overline{\delta}$   and $\overline{\delta_0}$ generate strongly continuous one-parameter groups of $*$-automorphisms $\alpha$ and $\alpha_0$ of $\cU ,$ such that 
\begin{equation*}
\lim_{\Lambda\nearrow\allspace}\|\alpha^t (A)-e^{t\delta_\Lambda}(A)\| =0,
\end{equation*}
and 
\begin{equation*}
\lim_{\Lambda\nearrow\allspace}\|\alpha_0^t (A)-e^{t\delta_{0,\Lambda}}(A)\| =0,
\end{equation*}
for all $A\in \cU,$ uniformly in $t\in {\mathbb R}$ for $t$ in compacts.  
\end{lemma}

We assume that the initial state of the system, $\omega_0,$ is given by 
\begin{equation*}
\omega_0 (A):= \lim_{\Lambda\nearrow \allspace} \frac{Tr_{\cH_{\Lambda}}(e^{-\sum_{i=1}^2\beta_i H_{R_i}(\Lambda)}A)}{Tr_{\cH_{\Lambda}}e^{-\sum_{i=1}^2\beta_i H_{R_i}(\Lambda)}},
\end{equation*}
for $A\in \cU,$ where $\beta_i \ge 0.$ The existence of the initial state follows from assumption (A1); see for example \cite{BR,Ruelle}. We note that the initial state $\omega_0$ is $\alpha_0$-invariant on $\cU,$ and the initial state of ${\mathbf R}_i, i=1,2,$ is the equilibrium KMS state at inverse temperature $\beta_i.$  Furthermore, $\omega_0$ is modular.\footnote{We recall some basic definitions from operator algebras. Consider a state $\omega$ on a $C^*$-algebra $\cU.$ We denote by $(\cH_{\omega},\pi_{\omega}, \Omega_{\omega})$ the  GNS-representation of $\cU$ corresponding to the state $\omega,$ and by $\cM_{\omega}$ the envelopping von Neumann algebra $\pi_{\omega}(\cU)'' .$ A state $\omega$ is called modular if its extension to $\cM_\omega$ is faithful, $\ie,$ if $A\Omega_{\omega}=0$ for $A\in \cM_{\omega},$ then $A=0.$ A state $\eta$ on $\cU$ is said to be $\omega$-normal if $\exists \rho$ a density matrix on $\cH_\omega$ such that $\eta(\cdot)= Tr_{\cH_{\omega}}(\rho\pi_\omega(\cdot)).$ A state $\eta$ on $\cU$ is said to be $\omega$-singular if $\eta\ge \lambda\phi,$ for some nonnegative $\lambda$ and $\omega$-normal state $\phi,$ implies $\lambda=0.$ }

We make the following additional assumptions. 

\begin{itemize}

\item[(A3)] {\it Homogeneity of the interaction.} Denote by $\tau$ the automorphism acting on $\cup_{\Lambda\subset \allspace}\cU_\Lambda$ corresponding to translations. If $A\in \cU_\Lambda,$ then $\tau_x(A)\in \cU_{\Lambda+x}.$ We assume that $\Phi$ is translationally invariant, $\ie, $ for $\Lambda\subset \allspace$ and $x\in\allspace,$
\begin{equation*}
\Phi(\Lambda+x)= \Phi(\Lambda).
\end{equation*}

\item[(A4)] {\it Scattering endomorphisms.} The limits
\begin{equation*}
\gamma_{\pm} (A) := \lim_{t\rightarrow \pm \infty} \alpha_0^{-t} \circ \alpha^t (A) 
\end{equation*}
exist in norm for all $A\in \cU.$ Furthermore, $\gamma_{\pm}$ are norm-preserving $*$-morphisms of $\cU.$ 

\end{itemize}

We now recall the notion of a nonequilibrium steady state (NESS) which was first introduced in \cite{Ruelle1}. We define the NESS as 
\begin{equation}
\label{eq:NESS}
\omega_+:= \lim_{t\rightarrow \infty} \omega_0\circ\alpha^t = \omega_0\circ\gamma^+.
\end{equation} 
Note that by construction $\omega_+$ is an $\alpha-$invariant state on $\cU.$ Furthermore, since $\gamma_+$ is a norm-preserving $*$-morphism and $\omega_0$ is a positive functional, $\omega_+$ is a positive functional, $\ie, \omega_+(A^*A)\ge 0$ for all $A\in \cU.$ \footnote{One can show for specific models that $\omega_+$ is $\omega_0$-singular, and that the entropy production of the coupled system in NESS is strictly positive, see for example \cite{FMU03,JP1}.}

\begin{remarks*}

(1) Since the coupling $V$ is bounded, it follows from a Dyson series expansion that
\begin{equation*}
\omega_+(A)= \omega_0(A) + \sum_{m\ge 1}(i)^m \int_0^\infty dt_1 \int_0^{t_1} dt_2 \cdots \int_0^{t_{m-1}} dt_m \omega_0([\alpha_0^{t_m}(V), [ \cdots [\alpha_0^{t_1}(V), A]]]),
\end{equation*}
for all $A\in \cU.$ This expansion is useful in computing the rates of heat transfer and entropy production; see for example \cite{Ruelle2, FMU03,JP1}.

(2) We note that (A4) follows for example from the assumption that there exists a norm-dense subalgebra $\cU_0\subset \cU$ such that the $C^*$-dynamical systems $(\cU,\alpha)$ and $(\cU,\alpha_0)$ are $L^1(\cU_0)$- asymptotically abelian, $\ie,$ 
\begin{eqnarray*}
\int_{-\infty}^\infty dt \| [A,\alpha^t (B)]\| &<& \infty, \\
\int_{-\infty}^\infty dt \| [A,\alpha_0^t (B)]\| &<& \infty
\end{eqnarray*}
for all $A,B\in \cU_0.$ \cite{Ruelle1} 
The assumption of asymptotic abelianess implies a form of ergodicity of the system, and it has been verified in only few physical models, such as the ideal Fermi gas.\cite{FMU03,BM83}

(3) One can relax assumption (A4) by weakening the definition of the NESS; (see for example \cite{Ruelle2,JP1}). Suppose only assumptions (A1) and (A2) hold. Then Lemma \ref{lm:Evolution} follows. We define the NESS as the limit  
\begin{equation*}
\omega_+ := w^*-\lim_{T\rightarrow\infty} \frac{1}{T}\int_0^T dt \omega_0\circ\alpha^t ,
\end{equation*}
in the weak-* topology, $\ie, \exists$ a sequence $\{T_n\}$ such that $\lim_{n\rightarrow\infty}T_n=\infty$ and  
\begin{equation*}
\omega_+ = \lim_{n\rightarrow \infty} \frac{1}{T_n}\int_0^{T_n} dt \omega_0\circ\alpha^t .
\end{equation*}
The dual $\cU^*$ of $\cU$ is compact in the weak-$*$ topology, and hence the set of all NESS is nonempty. If assumption (A4) holds, then $\omega_+= \omega_0\circ\gamma_+$ is unique. 

(4) Assumption (A3) is a technical one, and it simplifies the application of the cluster expansion method. 

\end{remarks*}

\subsection{Description of the Problem}

We now describe the problem we are interested in. Given a self-adjoint element $A\in\cU,  A^*=A,$ we introduce the probability distribution  $\sP_A,$ which, for $O$ a Borel measurable subset of ${\mathbb R},$ is given by
\begin{equation}
\label{eq:ProbMeasure}
\sP_A(O)=\omega_+({\mathcal I}_O(A)),
\end{equation} 
where $\omega_+$ is the NESS defined in (\ref{eq:NESS}) and ${\mathcal I}_O$ is the indicator function of $O\subset{\mathbb R}.$ Note that $\sP_A(O)$ denotes the probability that the expectation value of $A$ in state $\omega_+$ takes values in $O.$ It follows from (\ref{eq:ProbMeasure}) and the spectral theorem that, for a Borel measurable function $f,$ 
\begin{equation*}
\int_{\sigma(A)} \sP_A(dx) f(x) = \omega_+(f(A)),
\end{equation*}
where $\sigma(A)\subset {\mathbb R}$ is the spectrum of $A.$  

For every $X\in\allspace,$ we associate $A(X)\in \cU_X^i,$ such that $A(X)$ is selfadjoint. For $\Lambda\subset\allspace,$ we define the local observable $A_\Lambda:= \sum_{X\subset\Lambda} A(X).$ We also define
\begin{equation*}
\overline{A}_\Lambda:= \frac{1}{|\Lambda|}\sum_{X\in\Lambda}A(X).
\end{equation*}
It follows that $\sP_{\oA_\Lambda}$ is the probability measure associated to the empirical measure of $A.$  
We denote the fluctuation of $\oA_\Lambda$ by 
\begin{equation*}
\delta \oA_\Lambda := \oA_\Lambda - \omega_+(\oA_\Lambda).
\end{equation*} 
We ask whether 
\begin{equation}
\label{eq:Fluctuation}
W_{\Lambda}:= \frac{1}{\sqrt{|\Lambda|}} \sum_{x\in\Lambda} (A_x-\omega_+(\oA_\Lambda)) = \sqrt{|\Lambda|}\delta \oA_\Lambda
\end{equation}
obeys the {\it central limit theorem}, $\ie, \exists \sigma^2\ge 0$ such that, for all $z\in {\mathbb R},$
\begin{equation}
\lim_{\Lambda\nearrow \allspace}\omega_+ (e^{izW_{\Lambda}})=e^{-z^2 \sigma^2/2}, \label{eq:CLT}
\end{equation}
where 
\begin{equation*}
\sigma^2=\lim_{\Lambda\nearrow \allspace}|\Lambda| \omega_+ ((\delta \oA_\Lambda)^2).
\end{equation*}
Another question that we address is whether the sequence 
\begin{equation*}
\{ \sP_{\oA_\Lambda}: \Lambda \subset \allspace\} 
\end{equation*}
satisfies the {\it large deviation principle}, $\ie,$ whether there exists a lower semi-continuous convex function $I:{\mathbb R}\rightarrow {\mathbb R},$ the so called rate or entropy function, such that
\begin{eqnarray}
\limsup_{\Lambda\nearrow \allspace} \frac{1}{|\Lambda|} \log \sP_{\oA_\Lambda} (C) &\le& -\inf_{x\in C} I(x), \; C\subset {\mathbb R} \; closed \label{eq:LDP1} \\
\liminf_{\Lambda\nearrow \allspace} \frac{1}{|\Lambda|} \log \sP_{\oA_\Lambda} (O) &\ge& -\inf_{x\in O} I(x), \; O\subset {\mathbb R} \; open .  \label{eq:LDP2}
\end{eqnarray}
We now state the following assumption on the observable.  
 
 
\begin{itemize}

\item[(A5)] {\it Asymptotic evolution of the fluctuation.} $$s-\lim_{t\rightarrow\infty} \alpha^t(\oA_\Lambda)=\oA'_\Lambda + R_\Lambda,$$
where $\oA'_\Lambda = \frac{1}{|\Lambda|}\sum_{\Lambda'\subset\Lambda}A'(X),$ $A': \Lambda\rightarrow \cU$ is selfadjoint such that there exists $\lambda'>0$ with 
\begin{equation*}
\|A'\|_{\lambda'} = \sup_{x\in\allspace} \sum_{n\ge 0} e^{n\lambda'} \sum_{\stackrel{X\ni x}{|X|=n+1}} \|A'(X)\| <\infty,
\end{equation*}
and there exists a finite constant $C$ independent of $\Lambda$ such that $$\lim_{\Lambda\nearrow\allspace}|\Lambda | \|R_\Lambda-C\| = 0.$$ 
\end{itemize}

Assumption (A5) is trivially satisfied if $A$ commutes with the generator of the time evolution of the coupled system, $\ie,$ the observable is a conserved quantity of the coupled system, in which case $R_\Lambda=0$ and $A'=A.$ We remark how this assumption can be relaxed after Theorem \ref{thrm:Main}, Section \ref{sec:Main}.


\section{Main Result}\label{sec:Main}

In proving the central limit theorem and the large deviation principle, it is often useful to introduce the moment generating function, which physicists call {\it``free energy''}. We denote by $F$ the moment generating function, which is given by
\begin{equation}
\label{eq:MGfunction}
F(a):= \lim_{\Lambda\nearrow \allspace} \frac{1}{|\Lambda|} \log \omega_+ (e^{a |\Lambda| \oA_\Lambda}) = \lim_{\Lambda\nearrow \allspace} \frac{1}{|\Lambda|} \log \omega_+ (e^{a\sum_{X\in\Lambda} A(X)}).
\end{equation}

If $F$ is analytic is a neighborhood of the origin, Proposition 1 in \cite{Bryc93} imply the nonequilibrium central limit theorem (\ref{eq:CLT}).
Furthermore, if $F$ exists and is in $C^1({\mathbb R}),$ then, by the Ellis-G\"artner theorem\footnote{see for example \cite{DZ98}}, the large deviation principle holds and the rate (or entropy) function is 
\begin{equation}
\label{eq:Rate}
I(x)=\sup_{a\in {\mathbb R}}(a x -F(a)),
\end{equation}
the Legendre transform of $F.$
We note that if $F$ exists without being differentiable, then the least upper bound (\ref{eq:LDP1}) holds, but the lower bound (\ref{eq:LDP2}) fails  (see \cite{DZ98} for further details).  We now state the main result.

\begin{theorem}\label{thrm:Main}

(i)Suppose assumptions (A1)-(A5) hold. Then there exists positive constants $\tilde{\beta}$ and $\tilde{\mu},$ which depend on $\lambda$ and $\lambda'$ appearing in (A1) and (A5) respectively, such that, for $|\beta_{1,2}|<\tilde{\beta},$ the moment generating function $F$ defined in (\ref{eq:MGfunction}) is analytic in a for $a\in {\mathcal N}_{\tilde{\mu}}:= \{z\in {\mathbb C}: |z|\le \tilde{\mu}\},$ and $W_\Lambda$ defined in (\ref{eq:Fluctuation}) satisfies the nonequilibrium central limit theorem (\ref{eq:CLT}).

(ii) If, in addition, $A'$ appearing in (A5) is onsite, then, for $|\beta_{1,2}|<\tB,$ the moment generating function $F$ exists and is analytic in the strip $I_{\frac{\log 2}{\|A'\|}}:=\{ z\in {\mathbb C}: |\Im z |<\frac{\log 2}{\|A'\|}\},$ and the large deviation principle (\ref{eq:LDP1}) and (\ref{eq:LDP2}) hold, where the rate function $I$ is given by (\ref{eq:Rate}).

\end{theorem}

We prove both claims in the following section. 

\begin{remarks*}

\begin{itemize}

\item[(1)] In the particular case when the interaction between the reservoirs is turned off, each reservoir is at equilibrium, and claim (i) implies that the fluctuations of general observables in each reservoir at equilibrium satisfy a central limit theorem. This is a result which has been announced in \cite{NR04} and \cite{LR-B05}, but for which no explicit proof has been provided so far.  On the other hand, claim (ii) is a straightforward extension of the results of \cite{NR04} to the case at hand; see also \cite{LR-B05}.

\item[(2)] One may relax assumption (A5). Under the weaker assumption
\begin{equation*}
s-\lim_{t\rightarrow\infty}\alpha^t(\oA_{\Lambda\backslash\Lambda_{int}}) = \oA_{\Lambda\backslash\Lambda_{int}'}' + R_\Lambda,
\end{equation*}
where $\Lambda_{int}$ and $\Lambda_{int}'$ are {\it fixed} bounded regions of $\allspace$ containing the interaction region $\Lambda_0,$ and $\oA'_\Lambda$ and $R_\Lambda$ satisfy the conditions of assumption (A5), one can show that
\begin{equation*}
F'(a)=\lim_{\Lambda\nearrow\allspace} \frac{1}{|\Lambda\backslash\Lambda_{int}|} \log\omega_+ (e^{a|\Lambda\backslash\Lambda_{int}|\oA_{\Lambda\backslash\Lambda_{int}}})
\end{equation*}
is analytic in $a$ in some neighbourhood of the origin. (This follows from claim (i) in Theorem \ref{thrm:Main} and the fact that $\Lambda\nearrow\allspace$ in the van Hove limit; see proof of (i) in the following section.) Therefore, fluctuations of the observable in $\Lambda_{int}^c= \allspace\backslash\Lambda_{int}$ obey a central limit theorem. Furthermore, claim (ii) holds in the particular case when $A'$ is onsite, which implies that the large deviation principle holds for fluctuations in $\Lambda_{int}^c$. The latter is satisfied, for example, when $A$ is onsite and commutes with the generator of the free evolution, $\delta_0(A_x)=0,x\in\allspace,$ and the interaction $\Psi$ has a finite range $r.$ In this case, $A_x$ commutes with the coupling $V$ for $x\in \Lambda_{int}=\{ x\in\allspace: dist(x,\Lambda_0)>r \},$ and hence $\delta(A_x)=\delta_0(A_x)+i[V,A_x]=0,$ $\Lambda_{int}'=\Lambda_{int}, A=A',$ and $R_\Lambda=0.$     

\item[(3)] One can prove similar results if one has specific additional information about the correlations of the fluctuations, such as in systems which can be mapped onto quasi-free nonequilibrium system. Suppose for instance that there exists a constant $k>0$ independent of $\Lambda\subset \allspace, $ such that the correlation function of the fluctuation satisfy
\begin{equation*}
|\omega_+((\delta \oA_{\Lambda})^n)| \le n! \frac{k^n}{|\Lambda|^{n-1}} ,
\end{equation*}
for all $n> 1.$
In this case, one can easily show that $F$ is analytic in $a$ for $|a|<k,$ and hence the fluctuations $W_\Lambda$ defined in (\ref{eq:Fluctuation}) satisfy the nonequilibrium central limit theorem (\ref{eq:CLT}). 
\end{itemize}

\end{remarks*}



\section{Proof of the Main Result}\label{sec:Proof}

\subsection{Cluster Expansion}

We start by proving a useful lemma. It follows from assumptions (A1) and (A5) that there exist positive constants $\tB,\tA$ and $\tilde{\mu},$ which depend on $\lambda$ and $\lambda',$ such that
\begin{equation}
\label{eq:DecEst1}
\sup_{x\in{\mathbb Z}^d} \sum_{\Lambda\ni x} e^{2\tA|\Lambda|} (e^{\tilde{\mu} \|A'(\Lambda)\|+\tB \|\Phi^1 (\Lambda)+ \Phi^2 (\Lambda)\|}-1)\le \tA.
\end{equation}
For $\Lambda\subset {\mathbb Z}^d,$ we define 
\begin{equation}
\label{eq:Xi}
\xi (\Lambda):= e^{\tA |\Lambda |} (e^{\tilde{\mu} \|A' (\Lambda)\| + \tB \|\Phi^1 (\Lambda)+\Phi^2(\Lambda)\|}-1),
\end{equation}
and $\xi(\phi)=1.$
We denote by $\cS$ the set of finite collections of subsets of ${\mathbb Z}^d.$ For $S= \{\Lambda_1,\cdots,\Lambda_k\}\in \cS,$ we let $|S|=k,$ the cardinality of $S.$ We define the function
\begin{equation}
\label{eq:Theta}
\Theta(S|\Lambda) := \sum_{\nu\ge 1}^{|S|} \sum_{\Lambda\in \{\Lambda_1,\cdots ,\Lambda_\nu \}_{c} \subset S } \prod_{\mu=1}^{\nu} \xi(\Lambda_\mu) , 
\end{equation}
for $S\in \cS$ and $\Lambda\in S,$ where $\{\cdot\}_c$ denotes the collection of connected subsets of $\allspace.$ For $S\in \cS,$ we define $\underline{S}:= \cup_{i=1}^{|S|} \Lambda_i.$ We have the following lemma. 

\begin{lemma}\label{lm:DecayEstimate}
Suppose assumptions (A1) and (A5) hold. Then 
\begin{equation}
\label{eq:DecEst2}
\sup_{\stackrel{S\in \cS}{x\in \underline{S}}} \sum_{\stackrel{\Lambda\ni x}{\Lambda\in S}} \Theta (S|\Lambda) \le \tA,
\end{equation}
where $\tA$ appears in (\ref{eq:DecEst1}) and $\Theta$ is defined in (\ref{eq:Theta}). 
\end{lemma}

\begin{proof}[Proof of Lemma \ref{lm:DecayEstimate}.]
The proof is based on induction in the cardinality of $S\in\cS.$ The case $|S|=1$ follows trivially from (\ref{eq:DecEst1})-(\ref{eq:Theta}). Suppose that 
\begin{equation}
\label{eq:BaseStep}
\sup_{\stackrel{S\in \cS, |S|=N}{x\in \underline{S}}}\sum_{\stackrel{\Lambda\ni x}{\Lambda\in S}} \Theta(S|\Lambda)\le \tA.
\end{equation}
Now, for $|S|=N+1,$
\begin{align*}
\sum_{\stackrel{\Lambda\ni x}{\Lambda\in S}} \Theta(S|\Lambda) &=\sum_{\stackrel{\Lambda\ni x}{\Lambda\in S}} \sum_{\nu=1}^{N+1} \sum_{\Lambda\in \{\Lambda_1,\cdots ,\Lambda_\nu \}_{c} \subset S } \prod_{\mu=1}^{\nu} \xi(\Lambda_\mu) \\
&= \sum_{\stackrel{\Lambda\ni x}{\Lambda\in S}} \xi(\Lambda) \sum_{\nu=0}^{N} \sum_{\Lambda\in \{\Lambda_1,\cdots ,\Lambda_\nu \}_{c} \subset S\backslash \{\Lambda\} } \prod_{\mu=1}^{\nu} \xi(\Lambda_\mu) \\
&\le \sum_{\stackrel{\Lambda\ni x}{\Lambda\in S}} \xi(\Lambda) \sum_{\nu\ge 0}\frac{1}{\nu!} \sum_{r_1,\cdots ,r_\nu} \prod_{s=1}^\nu (\sum_{\stackrel{\{\Lambda_1,\cdots ,\Lambda_{r_s}\}_{c} \subset S\backslash \{\Lambda \} }{\Lambda \cap (\cup_{i=1}^{r_s}\Lambda_i)\ne \phi}} \prod_{i=1}^{r_s}\xi(\Lambda_i)) \\
&\le \sum_{\stackrel{\Lambda\ni x}{\Lambda\in S}} \xi(\Lambda) \sum_{\nu\ge 0}\frac{1}{\nu!} (\sum_{\stackrel{\Lambda'}{\Lambda\cap\Lambda' \ne \phi}} \xi(\Lambda') \sum_{r\ge 0} \sum_{\stackrel{\{\Lambda_1,\cdots,\Lambda_r \}\subset S\backslash \{\Lambda,\Lambda' \}}{\{\Lambda,\Lambda',\Lambda_1,\cdots,\Lambda_r\}_{c}}} \prod_{i=1}^r \xi(\Lambda_i) )^\nu \\
&\le \sum_{\stackrel{\Lambda\ni x}{\Lambda\in S}} \xi(\Lambda) e^{|\Lambda|\sup_{x'\in \underline{S\backslash \{\Lambda\}}}\sum_{\stackrel{\Lambda'\ni x'}{\Lambda'\in S\backslash \{\Lambda\}}}\Theta(S\backslash \{\Lambda\}|\Lambda')}.
\end{align*}
Together with (\ref{eq:BaseStep}), this implies
\begin{equation}
\label{eq:InducStep2}
\sup_{\stackrel{S\in \cS ,|S|=N+1}{x\in\underline{S}}}\sum_{\stackrel{\Lambda\ni x}{\Lambda\in S}} \Theta(S|\Lambda) \le \sum_{\stackrel{\Lambda\ni x}{\Lambda\in S}} \xi(\Lambda) e^{|\Lambda| \tA} .
\end{equation}
However, we know from (\ref{eq:DecEst1}) and (\ref{eq:Xi}) that 
\begin{equation*}
\sum_{\stackrel{\Lambda\ni x}{\Lambda\in S}} \xi(\Lambda) e^{|\Lambda| \tA} \le \sum_{\Lambda\ni x} e^{2\tA |\Lambda| } (e^{\tilde{\mu}\|A'(\Lambda)\|+\tB \|\Phi^1 (\Lambda)+\Phi^2(\Lambda)\|}-1) \le \tA. 
\end{equation*}
The last estimate together with (\ref{eq:InducStep2}) imply that 
\begin{equation*}
\sup_{\stackrel{S\in \cS,|S|=N+1}{x\in \underline{S}}}\sum_{\stackrel{\Lambda\ni x}{\Lambda\in S}} \Theta(S|\Lambda) \le \tA .
\end{equation*}
Claim (\ref{eq:DecEst2}) follows by induction.
\end{proof}

We now introduce further notation and definitions which are helpful in applying the cluster expansion method. For $\Lambda\subset\allspace,$
we consider $B$ a sequence of subsets of $\Lambda,$ whose union is connected,
\begin{equation*}
B=(\Lambda_1,\cdots,\Lambda_k)_c ,   
\end{equation*}
where the subscript $c$ stands for $ \cup_{i=1}^k \Lambda_i \subset\Lambda$ connected. 
We denote by $| B |$ the number of elements in the sequence $B, \ie, |B|=k,$ and let
\begin{equation*}
\underline{B}:= \cup_{r=1}^{| B |} \Lambda_r.
\end{equation*}
We denote by $\cC(\Lambda)$ the set of all such sequences of subsets in $\Lambda,$
\begin{equation*}
\cC(\Lambda):= \{ B=(\Lambda_1, \cdots, \Lambda_{|B|})_c : \Lambda\supset \underline{B}=\cup_{r=1}^{|B|}\Lambda_r  \; \; connected \} .
\end{equation*}
and we let
\begin{equation*}
\cC:=\cup_{\Lambda\nearrow\allspace}\cC(\Lambda).
\end{equation*} 
We introduce the measure $\mu_\Lambda$ on $\cC(\Lambda), \Lambda\subseteq\allspace,$ which, for $B\in \cC(\Lambda),$ is given by 
\begin{equation*}
\mu^a_\Lambda(B):= \frac{1}{dim(\cH_{\Lambda})}Tr_{\cH_{\Lambda}} \sum_{n=0}^{|B|} \prod_{r=1}^n (aA'(\Lambda_r)) \prod_{s=n+1}^{|B|}(-\beta_1 \Phi^1(\Lambda_s) - \beta_2\Phi^2(\Lambda_s)),
\end{equation*}
where $\Phi^i, i=1,2,$ is the interaction that appears in assumption (A1), $A'$ appears in assumption (A5), and $\Phi^i(\phi)=A'(\phi)=1.$ Furthermore, we denote $\mu^a:=\lim_{\Lambda\nearrow \allspace}\mu_\Lambda^a.$
Note that 
\begin{equation*}
|\mu^a_{\Lambda}(B)|\le 
(\sum_{n=0}^{|B|} \prod_{r=1}^n \|a A'(\Lambda_r) \|\prod_{s=n+1}^{|B|} \| \beta_1 \Phi^1(\Lambda_s)+\beta_2 \Phi^2(\Lambda_s)\|).
\end{equation*}
 We also introduce a real measurable function $\chi$ on $\cC(\Lambda) \times \cC(\Lambda)$ which, for $B_i$ and $B_j \in \cC(\Lambda),$ is given by
\begin{equation}  
\label{eq:Chi}
\chi(B_i,B_j)=
\begin{cases}
0 , \;  \underline{B_i}\cap\underline{B_j}=\phi \\
-1,  \; \; \; \; \underline{B_i}\cap\underline{B_j}\ne\phi
\end{cases}.
\end{equation}
The motivation for these choices will become apparent in Subsection \ref{subsec:pr:Main}. For $\Lambda\subset \allspace,$ we introduce the {\it partition function} 
\begin{equation}
\label{eq:AuxiliaryPF}
Z^a_\Lambda := 1+ \sum_{n\ge 1} \frac{1}{n!} \sum_{B_1,\cdots,B_n\in \cC(\Lambda)}\prod_{i=1}^n  \mu^a_\Lambda(B_i) \prod_{1\le i\le j\le n} (1+\chi(B_i,B_j)).
\end{equation}
We let $G_n$ be the set of all graphs with $n$ vertices, and $C_n\subset G_n$ the set of connected graphs with $n$ vertices. Given a graph $G,$ we denote by $E(G)$ the set of edges of $G$ and by $V(G)$ the set of vertices of $G.$ We introduce the combinatorial function $\varphi$ of finite sequences $(B_1,\cdots,B_n)$ of $\cC(\Lambda),\Lambda\subseteq\allspace,$ which is given by 
\begin{equation}
\label{eq:varphi}
\varphi (B_1,\cdots, B_n) := 
\begin{cases}
1 , n=1 \\
\frac{1}{n!} \sum_{G\in C_n} \prod_{(i,j)\in E(G)} \chi(B_i,B_j) , n\ge 2
\end{cases},
\end{equation}
where the product is over all edges of $G.$ If the graph with $n$ vertices and an edge between $i$ and $j$ whenever $\chi(B_i,B_j)\ne 0$ is connected, then the sequence $(B_1,\cdots, B_n)$ is connected. The cluster expansion allows for an expansion of $\frac{1}{|\Lambda|}\log Z^a_\Lambda$ in the limit $\Lambda\nearrow \allspace ; $ see for example \cite{Simon}. We will use the Kotecky-Preiss criteria to prove the convergence of the cluster expansion; see \cite{KP86,Ueltschi04}.  

\begin{proposition}\label{pr:ClusterExpansion}
Suppose assumptions (A1)-(A5) hold. Then, for $|a|\le \tilde{\mu}$ and $|\beta_{1,2}|<\tB,$ where $\tilde{\mu}$ and $\tB$ appear in (\ref{eq:Xi}),  
\begin{equation}
\label{eq:ClusterExpansion}
\lim_{\Lambda\nearrow\allspace} \frac{1}{|\Lambda|} \log Z^a_\Lambda = \sum_{n\ge 0} \sum_{\stackrel{B_1,\cdots, B_n\in\cC}{0\in\cup_{i=1}^n\underline{B_i}}}\frac{1}{|\cup_{i=1}^n \underline{B_i}|} \prod_{i=1}^n \mu^a(B_i) \varphi (B_1,\cdots,B_n) , 
\end{equation}
which is analytic in $a$ for $|a|\le \tilde{\mu}.$

\end{proposition}


\begin{proof}[Proof of Proposition \ref{pr:ClusterExpansion}]
We first show that for $\Lambda\subset\allspace, |a|\le \tilde{\mu}$ and $ |\beta_{1,2}|<\tB,$ 
\begin{equation}
\label{eq:ClusterEst1}
\sum_{\stackrel{B\in\cC(\Lambda)}{0\in\underline{B}}} |\mu^a_\Lambda(B)|e^{\tA |\underline{B}|} \le \tA,
\end{equation}
uniformly in $\Lambda.$ 
Let $S(B)$ be the image of $B\in\cC(\Lambda),$ $S(B)=\{\Lambda_1,\cdots,\Lambda_l\}_c,$ where $l\le |B|.$ Given $\Lambda_1,\cdots,\Lambda_n\subset\Lambda,$ such that $\cup_{i=1}^n\Lambda_i$ is connected, we have 
\begin{eqnarray*}
\sum_{\stackrel{B\in\cC(\Lambda)}{S(B)=\{\Lambda_1,\cdots,\Lambda_n\}_c}} |\mu^a_\Lambda(B)| 
&\le& \sum_{l\ge 1} \frac{1}{l!} \sum_{\stackrel{k_1,\cdots, k_n}{\sum_{i=1}^nk_i=l}} \frac{l!}{k_1!\cdots k_n!} \prod_{i=1}^n (\tilde{\mu}\|A'(\Lambda_i)\| + \tB\|\Phi^1(\Lambda_i)+\Phi^2(\Lambda_i)\|)^{k_i} \\
&=& \prod_{i=1}^n \sum_{k=1}^\infty \frac{1}{k!}(\tilde{\mu}\|A'(\Lambda_i)\|+\tB \|\Phi^1(\Lambda_i)+\Phi^2(\Lambda_i) \|)^k \\
&=& \prod_{i=1}^n (e^{\tilde{\mu}\|A'(\Lambda_i)\|+ \tilde{\beta} \|\Phi^1(\Lambda_i)+\Phi^2(\Lambda_i) \|}-1).
\end{eqnarray*}
Therefore,
\begin{equation*}
\sum_{\stackrel{B\in\cC(\Lambda)}{\underline{B}\ni x}} e^{\tA |\underline{B}|} \mu^a_\Lambda(B) = \sum_{n\ge 1} \sum_{\stackrel{\{\Lambda_1,\cdots,\Lambda_n\}_c}{x\in \cup_{i=1}^n\Lambda_i\subset\Lambda}} e^{\tA\sum_{i=1}^n|\Lambda_i|} \prod_{i=1}^n(e^{\tilde{\mu}\|A'(\Lambda_i)\|+\tB \|\Phi^1(\Lambda_i)+\Phi^2(\Lambda_i)\|}-1) ,
\end{equation*}
uniformly in $\Lambda\subset \allspace.$ Together with Lemma \ref{lm:DecayEstimate} and translational invariance, this implies (\ref{eq:ClusterEst1}). 
Furthermore, it follows from (\ref{eq:Chi}),(\ref{eq:ClusterEst1}) and translational invariance that 
\begin{equation}
\label{eq:ClusterEst2}
\sum_{B\in\cC(\Lambda)} |\mu^a_\Lambda(B)| |\chi(B,B')| e^{\tA|B|} = \sum_{\stackrel{B\in\cC(\Lambda)}{\underline{B}\cap\underline{B'}\ne\phi}} |\mu^a_\Lambda(B)| e^{\tA|B|} \le |\underline{B'}| \tA,
\end{equation}
uniformly in $\Lambda\in\allspace.$

Now, using (\ref{eq:ClusterEst1}) and (\ref{eq:ClusterEst2}), we inductively show that 
\begin{equation}
\label{eq:ClusterEst3}
1+\sum_{n=2}^N n\sum_{B_2,\cdots,B_n\in\cC(\Lambda)} \prod_{i=1}^n|\mu^a_\Lambda (B_i)| |\varphi(B_1,\cdots,B_n)|\le e^{\tA|\underline{B_1}|} ,
\end{equation}
for all $N\ge 2,$ uniformly in $\Lambda\subseteq\allspace .$
The case $N=2$ is trivially satisfied. Suppose that 
\begin{equation}
\label{eq:ClusterInductionStep}
1+\sum_{n=2}^{N-1} n\sum_{B_2,\cdots,B_n\in\cC(\Lambda)} \prod_{i=1}^n|\mu^a_\Lambda (B_i)| |\varphi(B_1,\cdots,B_n)|\le e^{\tA|\underline{B_1}|} ,
\end{equation}
uniformly in $\Lambda.$ It follows from (\ref{eq:varphi}) that
\begin{align*}
&\sum_{n=2}^N n\sum_{B_2,\cdots,B_n} \prod_{i=1}^n|\mu^a_\Lambda (B_i)| |\varphi(B_1,\cdots,B_n)| \\ &=  \sum_{n=2}^N \sum_{B_2,\cdots,B_n} \prod_{i=1}^n|\mu^a_\Lambda (B_i)| \frac{1}{(n-1)!} |\sum_{G\in C_n}\prod_{(i,j)\in E(G)}\chi(B_i,B_j)|.
\end{align*}
Let $G'$ be the graph one obtains after removing all the edges of $G$ with endpoints at 1, and let $(G_1,\cdots,G_n)$ be a sequence of connected graphs of $G'$ such that $\cup_{i=1}^kV(G_i)=\{2,\cdots,n\}$ and $V(G_i)\cap V(G_j)=\phi$ if $i\ne j. $ Note that to each graph $G'$ there corresponds $k!$ such sequences. Now,
\begin{equation*}
|\sum_{G\in C_n} \prod_{(i,j)\in E(G)} \chi(B_i,B_j)| \le \sum_{k\ge 1} \frac{1}{k!} |\sum_{G_1,\cdots, G_k} \prod_{l=1}^k [\prod_{(i,j)\in E(G_l)} \chi(B_i,B_j)\sum_{G_l'} \prod_{(i,j)\in E(G_l')} \chi(B_i,B_j)]|,
\end{equation*}  
where $G_l'$ runs over nonempty sets of edges with one endpoint at 1 and one in $V(G_l).$  Furthermore,
\begin{equation*}
|\sum_{G_l'} \prod_{(i,j)\in E(G_l')} \chi(B_i,B_j)| = |\prod_{i\in V(G_l)} (1+\chi(B_1,B_i))-1|\le \sum_{i\in V(G_l)} |\chi(B_1,B_i)|,
\end{equation*}
and hence 
\begin{align*}
1+&\sum_{n=2}^N n\sum_{B_2,\cdots,B_n} \prod_{i=1}^n|\mu^a_\Lambda (B_i)| |\varphi(B_1,\cdots,B_n)| \\ &\le   1+ \sum_{k\ge 1} \frac{1}{k!} |\sum_{G_1,\cdots, G_k} \prod_{l=1}^k [\prod_{(i,j)\in E(G_l)} \chi(B_i,B_j) \sum_{i\in V(G_l)}| \chi(B_1,B_i)|]|.
\end{align*}
We now perform the sum over the graphs $G_1,\cdots,G_k$ by summing over partitions of $\{2,\cdots,n\}$ in sets $V(G_i), i=1,\cdots, k,$ such that $\sum_{i=1}^k V(G_i)=n-1,$ and choosing connected graphs for each set of vertices. The number of such partitions is $\frac{(n-1)!}{\prod_{i=1}^k m_i!}.$ It follows that
\begin{align*}
1+&\sum_{n=2}^N n\sum_{B_2,\cdots,B_n} \prod_{i=1}^n|\mu^a_\Lambda (B_i)| |\varphi(B_1,\cdots,B_n)| \\ &\le 1 +\sum_{n=2}^N \frac{1}{k!} \sum_{\stackrel{m_1,\cdots,m_k\ge 1}{m_1+ \cdots + m_k=n-1}} \prod_{l=1}^k \sum_{B_1',\cdots, B_{m_l}'} \prod_{i=1}^{m_l} |\mu_\Lambda^a (B_i ')| |\varphi(B_1',\cdots, B_{m_l}')| \sum_{i=1}^{m_l}|\chi(B_1,B_i')|\\
&\le 1+\sum_{k\ge 1}\frac{1}{k!} \sum_{1\le m_1,\cdots, m_k\le N} \prod_{l=1}^k \sum_{B_1',\cdots,B_{m_l}'} \prod_{i=1}^{m_l} |\mu_\Lambda^a(B_i')||\varphi(B_1',\cdots,B_{m_l}')| \sum_{i=1}^{m_l} |\chi(B_1,B_i')| \\
&= 1 + \sum_{k\ge 1} \frac{1}{k!} [\sum_{m=1}^N \sum_{B_1',\cdots,B_m'} \prod_{i=1}^m |\mu_\Lambda^a(B_i')||\varphi(B_1',\cdots,B_m')| \sum_{i=1}^m |\chi(B_1,B_i')| ]^k \\
&\le 1 + \sum_{k\ge 1} \frac{1}{k!} \tA^k |\underline{B_1}|^k,
\end{align*} 
where we have used (\ref{eq:ClusterEst2}) and (\ref{eq:ClusterInductionStep}) in the last step. It follows by induction that (\ref{eq:ClusterEst3}) holds for all $N\ge 2.$ 

We now rewrite $Z_\Lambda^a$ as an exponential. Consider sequences of connected graphs $(G_1,\cdots,G_k)$ whose set of vertices form a partition of $\{1,\cdots,n\}.$ Summing over the number of vertices of each partition, and then over the partition, we get
\begin{align*}
Z^a_\Lambda &= 1 + \sum_{n\ge 1} \sum_{k\ge 1} \frac{1}{k!} \sum_{\stackrel{m_1,\cdots,m_k}{m_1+\cdots+m_k=n}} \frac{1}{\prod_{l=1}^k m_l!} \times\\
&\times \prod_{l=1}^k \{\sum_{B_1,\cdots, B_{m_l}} \prod_{i=1}^{m_l} \mu_\Lambda^a (B_i) \sum_{G\in C_{m_l}} \prod_{(i,j)\in E(G)} \chi(B_i,B_j) \}\\
&= 1+ \sum _{n\ge 1}\sum_{k\ge 1} \frac{1}{k!} \sum_{\stackrel{m_1,\cdots,m_k\ge 1}{m_1+\cdots +m_k=n}} \prod_{l=1}^k \sum_{B_1,\cdots,B_{m_l}} \prod_{i=1}^{m_l} \mu_\Lambda^a (B_i) \varphi(B_1,\cdots,B_{m_l}) \\
&= 1 + \sum_{k\ge 1} \frac{1}{k!} \{\sum_{n\ge 1} \sum_{B_1,\cdots,B_n} \prod_{i=1}^n \mu_\Lambda^a (B_i)\varphi(B_1,\cdots ,B_n)\}^k\\
&= e^{\sum_{n\ge 0} \sum_{B_1,\cdots,B_n} \prod_{i=1}^n \mu_\Lambda^a (B_i)\varphi(B_1,\cdots ,B_n)}.
\end{align*}
Note that together with (\ref{eq:ClusterEst2}), it follows that $|Z_\Lambda^a|\le e^{|\Lambda|\tA}.$ We now take the thermodynamic limit of well-defined quantities. Due to translational invariance,
\begin{align*}
&|\frac{1}{|\Lambda|} \log Z_\Lambda^a - \sum_{n\ge 0} \sum_{\stackrel{B_1,\cdots,B_n \in\cC(\Lambda) }{0\in \cup_{i=1}^n \underline{B_i}}} \frac{1}{|\cup_{i=1}^n\underline{B_i}|} \prod_{i=1}^n \mu_\Lambda^a (B_i)\varphi(B_1,\cdots ,B_n)|\\ &\le  \frac{1}{|\Lambda|}\sum_{x\in\Lambda} \sum_{n\ge 0} \sum_{\stackrel{B_1,\cdots,B_n}{x\in \cup_{i=1}^n \underline{B_i}\not\subset\Lambda}}  \prod_{i=1}^n |\mu^a (B_i)||\varphi(B_1,\cdots ,B_n)|.
\end{align*}
We now argue that the second term tends to zero as $\Lambda\nearrow \allspace$ in the van Hove sense; see for example \cite{BR}, Chapter 6. It follows from (\ref{eq:ClusterEst1}) and (\ref{eq:ClusterEst3}) that 
\begin{equation*}
\sum_{n\ge 0} \sum_{\stackrel{B_1,\cdots,B_n}{0\in \cup_{i=1}^n \underline{B_i}}} \prod_{i=1}^n |\mu_\Lambda^a (B_i)||\varphi(B_1,\cdots ,B_n)| <\tA,
\end{equation*}
uniformly in $\Lambda\in\allspace.$
For $\eta>0,\exists \Lambda'$ finite subset of $\allspace$ such that $$\sum_{n\ge 0} \sum_{\stackrel{B_1,\cdots,B_n}{0\in \cup_{i=1}^n \underline{B_i},\cup_{i=1}^n \underline{B_i}\not\subset \Lambda' }} \prod_{i=1}^n |\mu^a (B_i)||\varphi(B_1,\cdots ,B_n)|<\eta .$$ We let $$\Lambda_0':=\{x\in \Lambda : \Lambda'+x\subset\Lambda \}.$$ Then  
\begin{align*}
&\frac{1}{|\Lambda|}\sum_{x\in\Lambda} \sum_{n\ge 0} \sum_{\stackrel{B_1,\cdots,B_n}{x\in \cup_{i=1}^n \underline{B_i}\not\subset\Lambda}}  \prod_{i=1}^n |\mu^a (B_i)||\varphi(B_1,\cdots ,B_n)| \\ 
& = \frac{1}{|\Lambda|} (\sum_{x\in\Lambda_0'}+\sum_{x\in \Lambda\backslash\Lambda_0'}) \sum_{n\ge 0} \sum_{\stackrel{B_1,\cdots,B_n}{x\in \cup_{i=1}^n \underline{B_i}\not\subset\Lambda}}  \prod_{i=1}^n |\mu^a (B_i)| |\varphi(B_1,\cdots ,B_n)| \\
&\le \eta \frac{|\Lambda_0'|}{|\Lambda|}+ \tA \frac{1}{|\Lambda_0'|}. 
\end{align*}
Taking the limit $\eta\rightarrow 0$ and then $\Lambda\nearrow \allspace,$ the RHS of the last inequality tends to zero, and it follows that 
\begin{equation*}
\lim_{\Lambda\nearrow\allspace}\frac{1}{|\Lambda|} \log Z_\Lambda^a = \sum_{n\ge 0} \sum_{\stackrel{B_1,\cdots,B_n\in\cC}{0\in \cup_{i=1}^n \underline{B_i}}}  \frac{1}{|\cup_{i=1}^n \underline{B_i}|}\prod_{i=1}^n \mu^a (B_i)\varphi(B_1,\cdots ,B_n),
\end{equation*}
which is analytic in $a$ for $|a|\le \tilde{\mu}.$
\end{proof}


\subsection{Proof of Theorem \ref{thrm:Main}}\label{subsec:pr:Main}
\begin{proof}[Proof of (i)]

For $\Lambda\in\allspace,$ we define
\begin{equation*}
F_\Lambda (a) := \frac{1}{|\Lambda|} \log \frac{Tr_{\cH_{\Lambda}}(e^{a|\Lambda|\oA_\Lambda'}e^{-\beta_1H_{R_1}(\Lambda)-\beta_2 H_{R_2}(\Lambda)})}{Tr_{\cH_{\Lambda}} e^{-\beta_1 H_{R_1}(\Lambda)-\beta_2 H_{R_2}(\Lambda)}},
\end{equation*}
where $\oA_\Lambda'$ appears in assumption (A5). For $B=(\Lambda_1,\cdots,\Lambda_{|B|})_c\in \cC,$ we define 
\begin{equation*}
A'(B):= \prod_{i=1}^{|B|}aA'(\Lambda_i), \Phi^i(B):=\prod_{j=1}^{|B|} \beta_i\Phi^i(\Lambda_j),i=1=2,
\end{equation*}
and $A'(\phi)=\Phi^i(\phi)=1.$ 
We have
\begin{align*}
& e^{a|\Lambda|\oA_\Lambda'}e^{-\beta_1H_{R_1}(\Lambda)-\beta_2 H_{R_2}(\Lambda) } = \sum_{k\ge 0} \frac{1}{k!} (\sum_{X\subset\Lambda}a A'(X))^k \times \sum_{l\ge 0} \frac{1}{l!}(\sum_{X\subset \Lambda} -\beta_1 \Phi^1 (X)-\beta_2\Phi^2(X))^l \\
&= \{\sum_{k\ge 0} \frac{1}{k!} \sum_{n=1}^k  \sum_{\stackrel{B_1,\cdots,B_n\in \cC(\Lambda)}{\sum_{i=1}^n | B_i | =k}}\frac{(\sum_{i=1}^n | B_i |)!}{n!} \prod_{i=1}^n  A'(B_i)\prod_{1\le i\le j \le n} (1+\chi(B_i,B_j))\} \times\\ & \times  \{\sum_{l\ge 0} \frac{1}{l!}\sum_{j=1}^l  \sum_{\stackrel{B_1,\cdots,B_j\in \cC(\Lambda)}{\sum_{i=1}^j | B_i | =j}} \frac{(\sum_{i=1}^j | B_i |)!}{j!} \prod_{i=1}^j (-\Phi^1(B_i)-\Phi^2(B_i)) \prod_{1\le m\le n \le j} (1+\chi(B_m,B_n))\} .
\end{align*}
Grouping terms with common support, and using (\ref{eq:Chi}), one can rewrite the above expression as 
\begin{align*}
&e^{a|\Lambda|\oA_\Lambda'}e^{-\beta_1H_{R_1}(\Lambda)-\beta_2 H_{R_2}(\Lambda) } = \sum_{n=0}^\infty \frac{1}{n !} \sum_{B_1,\cdots,B_n\in \cC(\Lambda)} \prod_{m=1}^n  [\sum_{k\ge 0} \prod_{r=1}^k a A'(\Lambda_r) \prod_{s=1+k}^{|B_m|} ( -\beta_1 \Phi^1(\Lambda_s)- \\ &- \beta_2 \Phi^2(\Lambda_s))] \prod_{1\le i\le j \le \alpha} (1+\chi(B_i,B_j)) .
\end{align*}
Therefore,
\begin{equation*}
F_\Lambda (a) = \frac{1}{|\Lambda|} \{\log Z_\Lambda^a - \log Z_\Lambda^{a=0}\} ,
\end{equation*}
where $Z_\Lambda^a$ has been defined in (\ref{eq:AuxiliaryPF}). It follows from Proposition \ref{pr:ClusterExpansion} that 
\begin{equation}
\label{eq:FR1}
\lim_{\Lambda\nearrow\allspace} F_{\Lambda}(a) = \sum_{n\ge 0} \sum_{\stackrel{B_1,\cdots, B_n}{0\in\cup_{i=1}^n\underline{B_i}}}\frac{1}{|\cup_{i=1}^n \underline{B_i}|} (\prod_{i=1}^n \mu^a(B_i)-\prod_{i=1}^n \mu^0(B_i)) \varphi (B_1,\cdots,B_n),
\end{equation}
for $|\beta_{1,2}|\le \tB$ and $|a|\le \tilde{\mu},$ such that $\lim_{\Lambda\nearrow \allspace}F_{\Lambda}(a)$ is analytic in $a$ for $|a|\le \tilde{\mu}.$

We now define, for $\Lambda\subseteq \Lambda',$ 
\begin{equation*}
F_{\Lambda,\Lambda'}(a):= \frac{1}{|\Lambda|} \log \frac{Tr_{\cH_{\Lambda'}}(e^{a|\Lambda|\oA_\Lambda'}e^{-\beta_1H_{R_1}(\Lambda')-\beta_2 H_{R_2}(\Lambda')})}{Tr_{\cH_{\Lambda'}} e^{-\beta_1 H_{R_1}(\Lambda')-\beta_2 H_{R_2}(\Lambda')}}.
\end{equation*}

For $X,\Lambda\subset \allspace,$ we introduce the function 
\begin{equation*}
g(X,\Lambda) = 
\begin{cases}
1, \; \; X\subseteq \Lambda \\
0, \; \; otherwise
\end{cases},
\end{equation*}
and the measure on $\cC(\Lambda')$
\begin{equation}
\tilde{\mu}^a_{\Lambda,\Lambda'}(B):= \frac{1}{dim(\cH_{\Lambda'})}Tr_{\cH_{\Lambda'}} \sum_{n=0}^{|B|} \prod_{r=1}^n (aA'(\Lambda_r) g(\Lambda_r,\Lambda)) \prod_{s=n+1}^{|B|}(-\beta_1 \Phi^1(\Lambda_s) - \beta_2\Phi^2(\Lambda_s)),
\end{equation}
for $B\in \cC(\Lambda').$ We have 
\begin{align*}
|F_{\Lambda,\Lambda'}(a)-F_{\Lambda}(a)| &\le \frac{1}{|\Lambda|} \{ \sum_{n\ge 0}\sum_{\stackrel{B_1,\cdots,B_n\in \cC(\Lambda')}{\cup_{i=1}^n\underline{B_i}\cap\Lambda'\backslash\Lambda\ne\phi}} \prod_{i=1}^n |\tilde{\mu}^a_{\Lambda,\Lambda'}(B_i)| |\varphi(B_1,\cdots,B_n)| + \\
&+ \sum_{n\ge 0}\sum_{\stackrel{B_1,\cdots,B_n\in \cC(\Lambda')}{\cup_{i=1}^n\underline{B_i}\cap\Lambda'\backslash\Lambda\ne\phi}} \prod_{i=1}^n |\tilde{\mu}^{a=0}_{\Lambda,\Lambda'}(B_i)| |\varphi(B_1,\cdots,B_n)|\}.
\end{align*}
Since $\Lambda\nearrow \allspace$ in the van Hove sense, one can show using an argument similar to the one in the proof of Proposition \ref{pr:ClusterExpansion} that, for $|a|\le \tilde{\mu}$ and $|\beta_{1,2}|\le \tB,$
\begin{equation*}
\lim_{\Lambda\nearrow\allspace}\lim_{\Lambda'\nearrow\allspace} \frac{1}{|\Lambda|} \sum_{n\ge 0}\sum_{\stackrel{B_1,\cdots,B_n\in \cC(\Lambda')}{\cup_{i=1}^n \underline{B_i}\cap\Lambda'\backslash\Lambda\ne \phi}} \prod_{i=1}^n |\tilde{\mu}^a_{\Lambda,\Lambda'}(B_i)| |\varphi(B_1,\cdots,B_n)| = 0,
\end{equation*}
and hence
\begin{equation}
\label{eq:FEDiff}
\lim_{\Lambda\nearrow\allspace}\lim_{\Lambda'\nearrow\allspace} F_{\Lambda,\Lambda'}(a) = \sum_{n\ge 0} \sum_{\stackrel{B_1,\cdots, B_n}{0\in\cup_{i=1}^n\underline{B_i}}}\frac{1}{|\cup_{i=1}^n \underline{B_i}|} (\prod_{i=1}^n \mu^a(B_i)-\prod_{i=1}^n\mu^0(B_i)) \varphi (B_1,\cdots,B_n),
\end{equation}
which is analytic in $a$ for $|a|\le \tilde{\mu}.$

To complete the proof of claim (i), we {\it formally} rewrite the moment generating function $F$ using assumption (A5) as 
\begin{align*}
F(a)&= \lim_{\Lambda\nearrow\allspace} \lim_{t\rightarrow\infty} \frac{1}{|\Lambda|}\log \omega_0(\alpha^t(e^{a|\Lambda|\oA_\Lambda})) \\
&= \lim_{\Lambda\nearrow\allspace} \frac{1}{|\Lambda|} \log \omega_0(e^{a|\Lambda|\oA_{\Lambda}'+ |\Lambda| R_\Lambda} ) \\
&=\lim_{\Lambda\nearrow \allspace} \lim_{\Lambda'\nearrow\allspace} \frac{1}{|\Lambda |} \log ( \frac{Tr_{\cH_{\Lambda'}}e^{a \sum_{X\in\Lambda}A'(X) + |\Lambda|R_\Lambda} e^{-\sum_{i=1}^2 \beta_i H_{R_i}(\Lambda')}}{Tr_{\cH_{\Lambda'}}e^{-\sum_{i=1}^2 \beta_i H_{R_i}(\Lambda')}}) \\
&= \lim_{\Lambda\nearrow \allspace} \lim_{\Lambda'\nearrow\allspace} \frac{1}{|\Lambda |} \log ( \frac{Tr_{\cH_{\Lambda'}}e^{a \sum_{X\in\Lambda}A'(X)} e^{-\sum_{i=1}^2 \beta_i H_{R_i}(\Lambda')}}{Tr_{\cH_{\Lambda'}}e^{-\sum_{i=1}^2 \beta_i H_{R_i}(\Lambda')}}) + C,
\end{align*}
where the constant $C$ appears in assumption (A5). Together with (\ref{eq:FEDiff}), this implies that, for $|a|\le \tilde{\mu}$ and $|\beta_{1,2}|\le \tB$
\begin{equation*}
F(a)= \sum_{n\ge 0} \sum_{\stackrel{B_1,\cdots, B_n}{0\in\cup_{i=1}^n\underline{B_i}}}\frac{1}{|\cup_{i=1}^n \underline{B_i}|} (\prod_{i=1}^n \mu^a(B_i)-\prod_{i=1}^n \mu^0(B_i)) \varphi (B_1,\cdots,B_n) + C,
\end{equation*} 
which is analytic in $a.$ 
Claim (i) follows from analyticity of the moment generating function in a neighbourhood of the origin and Proposition 1 in \cite{Bryc93}.


\noindent{\it Proof of (ii).}

The proof of claim (ii) is a straight forward extension of the result in \cite{NR04}; see also \cite{LR-B05}. We sketch the main steps of the proof. We rewrite the moment generating function $F$ as 
\begin{align*}
F(a) =& \lim_{\Lambda\nearrow \allspace} \lim_{\Lambda'\nearrow\allspace} \frac{1}{|\Lambda |} \log ( \frac{Tr_{\cH_{\Lambda'}}e^{a \sum_{x\in\Lambda}A_x' + |\Lambda|R_\Lambda} e^{-\sum_{i=1}^2 \beta_i H_{R_i}(\Lambda')}}{Tr_{\cH_{\Lambda'}}e^{-\sum_{i=1}^2 \beta_i H_{R_i}(\Lambda')}}) \\
&= \lim_{\Lambda\nearrow \allspace} \lim_{\Lambda'\nearrow \allspace}  \frac{1}{|\Lambda |}\{\log \overline{Z}_{\Lambda,\Lambda'}^a 
-\log \overline{Z}_{\Lambda,\Lambda'}^{a=0}\} + C
\end{align*}
where 
\begin{equation*}
\label{eq:PartFunct2}
\overline{Z}_{\Lambda,\Lambda'}^a:= \frac{ Tr_{\cH_{\Lambda'}} e^{a \sum_{x\in\Lambda}A_x'}e^{-\sum_{i=1}^2 \beta_i H_{R_i}(\Lambda')}}{Tr_{\cH_{\Lambda'}}e^{a\sum_{x\in\Lambda}X_x}\otimes\unit_{\Lambda'\backslash\Lambda}} ,
\end{equation*}
and $\unit_{\Lambda'\backslash\Lambda}$ is the identity on $\Lambda'\backslash\Lambda.$
We define the measure on $\cC(\Lambda')$ 
\begin{equation*}
\overline{\mu}_{\Lambda,\Lambda'}^{a} (B):= \frac{Tr_{\cH_{\Lambda'}}(e^{a\sum_{x\in\Lambda}A_x'}\prod_{i=1}^{|B|} (-\beta_1\Phi^1(\Lambda_i)-\beta_2\Phi^2(\Lambda_i))}{Tr_{\cH_{\Lambda'}}(e^{a\sum_{x\in \Lambda}A_x'}\otimes\unit_{\Lambda'\backslash\Lambda})} ,
\end{equation*}
for $B\in \cC(\Lambda'),$ and $\overline{\mu}_{\Lambda,\Lambda'}^{a} (\phi)=1.$ 
We have 
\begin{equation}
\label{eq:PartFunct2}
\overline{Z}_{\Lambda,\Lambda'}^a = \sum_{n=0}^\infty \frac{1}{n !} \sum_{B_1,\cdots,B_n\in \cC(\Lambda')} \prod_{i=1}^n \overline{\mu}_{\Lambda,\Lambda'}^{a}(B_i)\prod_{1\le i\le j \le n} (1+\chi(B_i,B_j)). 
\end{equation}
Using the cluster expansion method, it is shown in \cite{NR04} that, for $|\beta_{1,2}|\le \tB$ and $a\in I_{\frac{\log 2}{\|A'\|}}:= \{z\in {\mathbb C}: |\Im z|< \frac{\log 2}{\|A'\|}\},$
\begin{equation*}
\lim_{\Lambda\nearrow\allspace} \lim_{\Lambda'\nearrow\allspace} \frac{1}{|\Lambda|} \log \overline{Z}_{\Lambda,\Lambda'}^a = \sum_{n\ge 0} \sum_{\stackrel{B_1,\cdots, B_n}{0\in\cup_{i=1}^n\underline{B_i}}}\frac{1}{|\cup_{i=1}^n \underline{B_i}|} \prod_{i=1}^n \overline{\mu}^a(B_i) \varphi (B_1,\cdots,B_n),
\end{equation*}
where $\overline{\mu}^a(B):= \lim_{\Lambda\nearrow\allspace}\lim_{\Lambda'\nearrow\allspace} \overline{\mu}_{\Lambda,\Lambda'}^a.$ 
This implies the analyticity of $F$ in $a$ for $a\in I_{\frac{\log 2}{\|A'\|}},$ which, together with the Ellis-G\"artner theorem, implies claim (ii).

\end{proof}

\bibliographystyle{amsplain}

\begin{thebibliography}{10}

\bibitem{Ellis05} R.S. Ellis, {\it Entropy, Large Deviations and Statistical Mechanics}, Springer (Classics in Mathematics), New York, 2005

\bibitem{DZ98} A. Dembo and O. Zeitouni, {\it Large Deviation Techniques and Applications}, Springer, 2nd edition, New York, 1998

\bibitem{GVV1} D. Goderis, A. Verbeure, and P. Vets, {\it Noncommutative central limits}, Prob. Th. Rel. Fields {\bf 82}, 527-544 (1989)

\bibitem{GVV2} D. Goderis, A. Verbeure, and P. Vets, {\it Dynamics of fluctuations for quantum lattice systems}, Commun. Math. Phys. {\bf 128}, 533-540 (1990)

\bibitem{M1} T. Matsui, {\it Bosonic central limit theorem for one-dimensional XY model}, Rev. Math. Phys. {\bf 14}, 675-700 (2002)

\bibitem{M2} T. Matsui, {\it On the algebra of fluctuations in quantum spin chains}, Ann. Henri Poincar\'e, {\bf 4},63-83 (2002)

\bibitem{LLS00} J.L. Lebowitz, M. Lenci and H. Spohn, {\it Large deviations for ideal quantum systems}, J. Math. Phys. {\bf 41}, 1224-1243 (2000)

\bibitem{GLM02} G. Gallavotti, J.L. Lebowitz and V. Mastropietro, {\it Large deviations for rarefied quantum gases}, J. Stat. Phys. {\bf 108}, 831-861 (2002)

\bibitem{NR04} K. Netocny and F. Redig, {\it Large deviations for quantum spin systems}, J. Stat. Phys. {\bf 117}, 521-547 (2004)

\bibitem{LR-B05}M. Lenci and L. Rey-Bellet, {\it Large deviations in quantum lattice systems: one-phase region}, J. Stat. Phys. {\bf 119}, 715-746 (2005)




\bibitem{DLS1} B. Derrida, J.L. Lebowitz and E.R. Speer, {\it Exact free energy functional for a driven diffusive open stationary nonequilibrium system}, Phys. Rev. Lett. {\bf 89}, 030601 (2002)

\bibitem{DLS2} B. Derrida, J.L. Lebowitz and E.R. Speer, {\it Large deviation of the density profile in the steady state of the open symmetric simple exclusion process}, J. Stat. Phys. {\bf 107}, 599-634 (2002)

\bibitem{BDGJL02} L. Bertini, A. De Sole, D. Gabrielli, G. Jona-Lasinio and C. Landim, {\it Macroscopic fluctuation theory for stationary nonequilibrium steady states}, J. Stat. Phys. {\bf 107}, 635-675 (2002)

\bibitem{BDGJL03} L. Bertini, A. De Sole, D. Gabrielli, G. Jona-Lasinio and C. Landim, {\it Large deviations for the boundary driven symmetric simple exclusion process}, Math. Phys. Analysis and Geometry {\bf 6}, 231-267 (2003)


\bibitem{Sewell06} G. Sewell, {\it Quantum macroscopic theory of nonequilibrium steady states}, Rev. Math. Phys. {\bf 17}, 977 (2005)

\bibitem{Bryc93} W. Bryc, {\it A remark on the connection between the large deviation principle and the central limit theorem}, Stat. Prob. Lett. {\bf 18}, 253-256 (1993)


\bibitem{BR} O. Bratteli and D. Robinson, {\it Operator Algebras and Quantum Statistical Mechanics}, Vol. $I \& II$, Springer, Berlin, 1987

\bibitem{Ruelle} D. Ruelle, {\it Statistical Mechanics: Rigorous Results}, World Scientific Publishing Company, 1999



\bibitem{Ruelle1} D. Ruelle, {\it Natural nonequilibrium states in quantum statistical mechanics}, J. Stat. Phys. {\bf 98}, 57-75 (2000)

\bibitem{Ruelle2} D. Ruelle, {\it Entropy production in quantum spin systems}, Commun. Math. Phys. {\bf 224}, 3-16 (2001)

\bibitem{FMU03} J. Fr\"ohlich, M. Merkli and D. Ueltschi, {\it Dissipative transport: thermal contacts and tunnelling junctions}, Ann. Henri Poincar\'e {\bf 4}, 897-945 (2003)

\bibitem{BM83} D.D. Botvich and V.A. Malyshev, {\it Unitary equivalence of temperature dynamics for locally perturbed Fermi gas}, Commun. Math. Phys. {\bf 91}, 301 (1983)


\bibitem{JP1} V. Jacksic and C.-A. Pillet, {\it Mathematical theory of non-equilibrium quantum statistical mechanics}, J. Stat. Phys {\bf 108}, 787, (2002)







\bibitem{Simon} B. Simon, {\it The Statistical Mechanics of Lattice Gases}, Princeton University Press (1993)

\bibitem{KP86} R. Kotecky and D. Preiss, {\it Cluster expansions for abstract polymer models}, Commun. Math. Phys. {\bf 103}, 491-498 (1986)

\bibitem{Ueltschi04} D. Ueltschi, {\it Cluster expansion and correlation functions}, Moscow Math. J. {\bf 4}, 511-522 (2004)






\end{thebibliography}

\end{document}